\documentclass[sigconf,author]{acmart}
\renewcommand\footnotetextcopyrightpermission[1]{}
\setcopyright{none}
\usepackage{tabularx}
\usepackage{float}
\usepackage{afterpage}
\usepackage{booktabs}
\usepackage{colortbl}
\usepackage{enumitem}
\usepackage{etoolbox}
\usepackage{graphicx}

\AtBeginDocument{%
  \providecommand\BibTeX{{%
    \normalfont B\kern-0.5em{\scshape i\kern-0.25em b}\kern-0.8em\TeX}}}

\begin{document}
\pagenumbering{arabic}

\title{A Survey of the Security Challenges and Requirements for \\ IoT Operating Systems}

\author{Alvi Jawad}
\email{alvi.jawad@carleton.ca}
\affiliation{%
  \institution{Department of Systems and Computer Engineering\\
  Carleton University}
  \streetaddress{1125 Colonel By Dr}
  \city{Ottawa}
  \state{Ontario}
  \postcode{K1S 5B6}
}


\begin{abstract}
The Internet of Things (IoT) is becoming an integral part of our modern lives as we converge towards a world surrounded by ubiquitous connectivity. The inherent complexity presented by the vast IoT ecosystem ends up in an insufficient understanding of individual system components and their interactions, leading to numerous security challenges. In order to create a secure IoT platform from the ground up, there is a need for a unifying operating system (OS) that can act as a cornerstone regulating the development of stable and secure solutions. In this paper, we present a classification of the security challenges stemming from the manifold aspects of IoT development. We also specify security requirements to direct the secure development of an unifying IoT OS to resolve many of those ensuing challenges. Survey of several modern IoT OSs confirm that while the developers of the OSs have taken many alternative approaches to implement security, we are far from engineering an adequately secure and unified architecture. More broadly, the study presented in this paper can help address the growing need for a secure and unified platform to base IoT development on and assure the safe, secure, and reliable operation of IoT in critical domains.

\end{abstract}


\maketitle

\section{Introduction}
The Internet of Things (IoT) defines the next generation of automated communication. IoT refers to an expanding interconnection of billions of embedded devices via the internet to enable information exchange and interaction among its components. The development and use of IoT devices are increasing at a staggering rate. According to a recent report from Gartner, the IoT industry is expected to grow to 5.8 billion enterprise and automotive endpoints in 2020\footnote{https://www.gartner.com/document/3955769}. However, this sudden proliferation also increases the risk associated with insecure IoT devices. The rapid-growth of exposed end-points and the inherent complexity of the network widens the attack surface, making the task of ensuring safety, security, and reliability for the entire system exceedingly difficult. The Mirai Botnet attack in 2016, affecting millions of poorly configured IoT devices to stage a massive distributed denial-of-service (DDoS) attack, is a prime example of this \cite{DDoS_Mirai}. A broader exploit could potentially lead to serious ramifications in safety-critical IoT systems such as the Internet of Medical Things (IoMT), autonomous vehicular networks, smart grids, and many more. Our inadequacy of system knowledge, coupled with increasingly resourceful adversaries, only intensifies this problem.
\par
Most IoT systems are open and scalable, meaning that the system components frequently participate in intensive information exchange within a growing environment. As the functionality of many devices is highly reliant on the data sent over the internet, ensuring the confidentiality, integrity, and availability of both stored and in-transit data becomes a top priority. This concern necessitates strong encryption algorithms and the need for mandatory end-to-end security \cite{Borgohain2015SurveyOS}. On the other hand, a significant number of issues can also result from the complex and poorly understood interactions among individual connected devices. If a user connects to a malicious or compromised device infected by a lurking malware, it could result in all the user's devices being contaminated as end-devices are rarely equipped with strong security defenses \cite{Objects_Taiwan}. Thus providing in-built security mechanisms for individual devices within the IoT system also becomes a necessity. 
\par
However, the objects (e.g., sensors, actuators, and other heterogeneous devices) connected to an IoT network are heavily constrained in terms of computing power, available memory, and energy capacities \cite{RIOT}. A typical example is wireless sensor networks (WSN), whose tiny sensor nodes require them to be extremely energy-efficient and work with very limited physical and logical resources \cite{TinyOS}. As a result, conventional means of securing a system, such as robust cryptographic algorithms, cannot be used as they heavily impact the performance, memory, and power consumption of these devices. Alongside these limitations, there are also inherent design and development challenges, and difficulties resulting from the heterogeneity and large scale development of IoT devices.
\par
The development of a secure operating system (OS) can resolve many of these challenges and direct the secure development of individual devices and applications. Typical OSs have minimum processing requirements that often exceed the capabilities of resource-starved IoT devices. As such, lightweight OSs specifically designed to work within scarce resource constraints, such as the Tiny operating system (TinyOS) and Contiki, have recently garnered much research attention in WSN \cite{Tkunz}. However, the studies and development have largely focused on the feasibility of running on resource-starved devices and less on developing secure implementations. This also underscores the importance of proper formulation, implementation, and enforcement of standard security requirements to guide the evolution of IoT OSs \cite{Denmark_Sec_Framework}.
\par
The main contributions of this paper are as follows.
\begin{itemize}
    \item Identification of the existing and emerging security challenges in IoT and classifying them into common and IoT-specific unique challenges.
    \item Determination of the role of an OS to resolve many of these challenges
    \item Defining specific security requirements to guide the secure development of an IoT OS 
    \item Security-centric assessment of some of the most prominent OSs in the IoT domain by comparing and contrasting their approach to security with the security requirements, and discussing security evaluations done on them.
\end{itemize}

This knowledge will help reinforce our understanding of deficiencies in the currently available IoT OSs and the risks of using them in safety-critical IoT systems. We will also gain insights into requirements to be considered when developing a unified OS for IoT, and how the security development of the current OSs should progress. It is important to note that this research was done as part of a graduate course and is far from exhaustive. The author is no longer working on it and is sharing the findings with others in hopes of benefitting someone out there.

The rest of this paper is organized as follows. Section 2 briefly discusses the terminologies and concepts used for the remainder of the paper. Section 3 explores and classifies common and unique security challenges. Section 4 follows from the previous discussion and examines the role of an OS in facilitating secure IoT development. Section 5 defines specific security requirements for architecting an IoT OS. Section 6 surveys some of the dominant IoT OSs and relevant security-centric evaluations. Finally, section 7 details related work in the literature, and section 8 concludes and describes our envisioned outline for future work.




\section{Background}
In this section, we discuss some terminologies and concepts used throughout this paper.

\subsection{Classes of Constrained Devices}
The diverse resource capabilities of IoT devices mandate the need to determine an area of focus to develop targeted security solutions. According to the Internet Engineering Task Force (IETF) classification \cite{IETF_Classes}, the three subcategories of constrained IoT devices are illustrated in table \ref{tab:Classes}.

\begin{center}
\begin{table}[h]
\caption{Classification of Constrained devices.}
\begin{tabularx}{0.475\textwidth}{ 
  | >{\centering\arraybackslash}X 
  | >{\centering\arraybackslash}X 
  | >{\centering\arraybackslash}X | }
 \hline
 Name & Data Size & Code Size \\
 \hline
 Class 0, C0  & $<<$ $10$ KiB  & $<<$ $100$ KiB \\
 
 Class 1, C1  & $\sim$ $10$ KiB  & $\sim$ $100$ KiB \\
 
 Class 2, C2 & $\sim$ $50$ KiB  & $\sim$ $250$ KiB \\
 \hline
\end{tabularx}
\label{tab:Classes}
\end{table}
\end{center}

Class 0 devices are the most resource-starved (e.g., customized sensor-like nodes having \(<\)\(<\)10 KiB of RAM and \(<\)\(<\)100 KiB of flash memory) among all IoT devices and often require the assistance of larger devices to act as gateways or proxies to participate in internet communications. These extreme resource constraints prohibit the implementation of any rigorous security mechanism, such as strong cryptographic algorithms. These devices are often minimally secured by an initial configuration and rarely reconfigured during their lifetime.

Class 1 devices are less constrained and can be implemented as peers into an IP network with these limitations in mind. However, they still cannot make use of a traditional full protocol stack, such as HTTP, transport
layer security (TLS), and related security protocols. These devices are capable of using lightweight protocols specifically designed for constrained devices (e.g., constrained application protocol (CoAP)) and does not require the need of a gateway to perform meaningful conversations with other nodes. 

Class 2 devices and beyond are much less constrained and typically capable of supporting most of the protocol stacks used on notebooks and servers. However, they can still benefit from using lightweight and energy-efficient implementations, reducing bandwidth consumption and development cost while increasing interoperability and the available resources to run applications.

Due to the extreme resource constraints, software developed for class 0 devices is often bare-metal and very hardware-specific \cite{HAL_IoT_OS_survey}. The development of an OS for these devices also emphasizes task specialization, and security concerns are often ignored to attain maximum device lifetime. Therefore, our analysis, meant to facilitate the development of a secure and unifying OS for all IoT devices, does not include class 0 devices. Further mentions of resource-starved devices in this paper will always refer to class 1 and class 2 devices unless otherwise mentioned.

\subsection{Security Properties}
In an IoT ecosystem, the inhabiting systems and components engage in frequent internal and external information exchange with the surrounding environment. The most significant security challenge faced by the IoT is the protection of the enormous amount of data that they store, use, or transmit. In this section, we briefly outline the security properties used throughout this paper, the preservation of which is integral to building an adequately secure and resilient IoT architecture.
\thispagestyle{empty}

\subsubsection{\textbf{Confidentiality and Privacy}}
Data collected and shared between devices may contain a large amount of private information to provide better services and fulfill personal preferences \cite{Objects_Taiwan}. Many applications, especially those deployed in the healthcare sector, generate traceable signatures of the location and the behavior of users. An unauthorized entity gaining access to a user device's stored and transmitted data can analyze the data to jeopardize user privacy. For example, a compromised security camera can give the attacker information about when a house or industrial location is occupied and when it is not and act as an exposed entry point to more vulnerable devices connected to the same network. This can engender serious confidentiality and privacy concerns, and thus personal and confidential information misuse must be prevented. Other examples of attacks that violate these properties include eavesdropping, side-channel attacks, traffic analysis, cryptanalysis attacks, etc. \cite{Denmark_Sec_Framework}.

\subsubsection{\textbf{Integrity and Authenticity}}
IoT represents the harmony of the digital and the physical world, where an attack that manipulates information on the internet can lead to controlling actuation in the physical world \cite{Cisco_white_paper}. Safety-critical devices (e.g., health monitoring devices, vehicular sensors, smart meters) depend heavily on the correctness and accuracy of the received data to make near-real-time decisions. The critical decisions are made based on the assumption that the data collected has not been manipulated in transit, and failure to prevent the unauthorized modification and corruption of the transmitted data can lead to aberrant and unwanted behaviors \cite{EWSN_SecAnalContiki}.

For proper scaling for IoT development, trust on the devices, and the architecture that it runs on will be a fundamental issue. Without proper authentication mechanisms in place, an adversary can masquerade as a legitimate entity within the network and use spoofed or compromised nodes to affect the operation and performance of the system. Therefore, protecting the integrity of both stored and in-transit data and validating its authenticity becomes a pressing issue as lack of countermeasures can cause severe damage to critical infrastructures such as health sectors and smart grids, and can even lead to loss of human lives. Attacks against the integrity and authenticity properties include physical and remote access tampering, spoofing, message modification and corruption, node subversion, and routing attacks, among others \cite{Denmark_Sec_Framework}.

\subsubsection{\textbf{Availability and Resiliency}}
The connected nature of the IoT infrastructure leaves it highly exposed to a range of network attacks. Unlike traditional information technology (IT) systems, critical IoT infrastructure (e.g., smart grid, Industrial IoT (IIoT)) cannot go through a complete system shutdown in the event of an incident, making availability a top priority \cite{hitachi_white_paper}. A critical industrial process may be highly reliant on the accurate and timely measurement of temperature, whereas a targeted DDoS attack can make resources unavailable to the endpoint \cite{Cisco_white_paper}. Furthermore, the compromise of a single endpoint can lead to subsequent exploitation of other devices on the network, potentially seizing control of critical operations. Therefore, we need concrete resiliency countermeasures to protect the critical functionalities of the system to allow critical operations to continue even when some parts of the system are compromised. The issues worsen as devices previously without connectivity become increasingly connected to the IoT, inherit the old and new vulnerabilities, and enlarge the attack surface even further.
\thispagestyle{empty}

\section{Challenges for dealing with the Security vulnerabilities in IoT}
\begin{figure*}[t]
    \centering
    \includegraphics[width=1\textwidth]{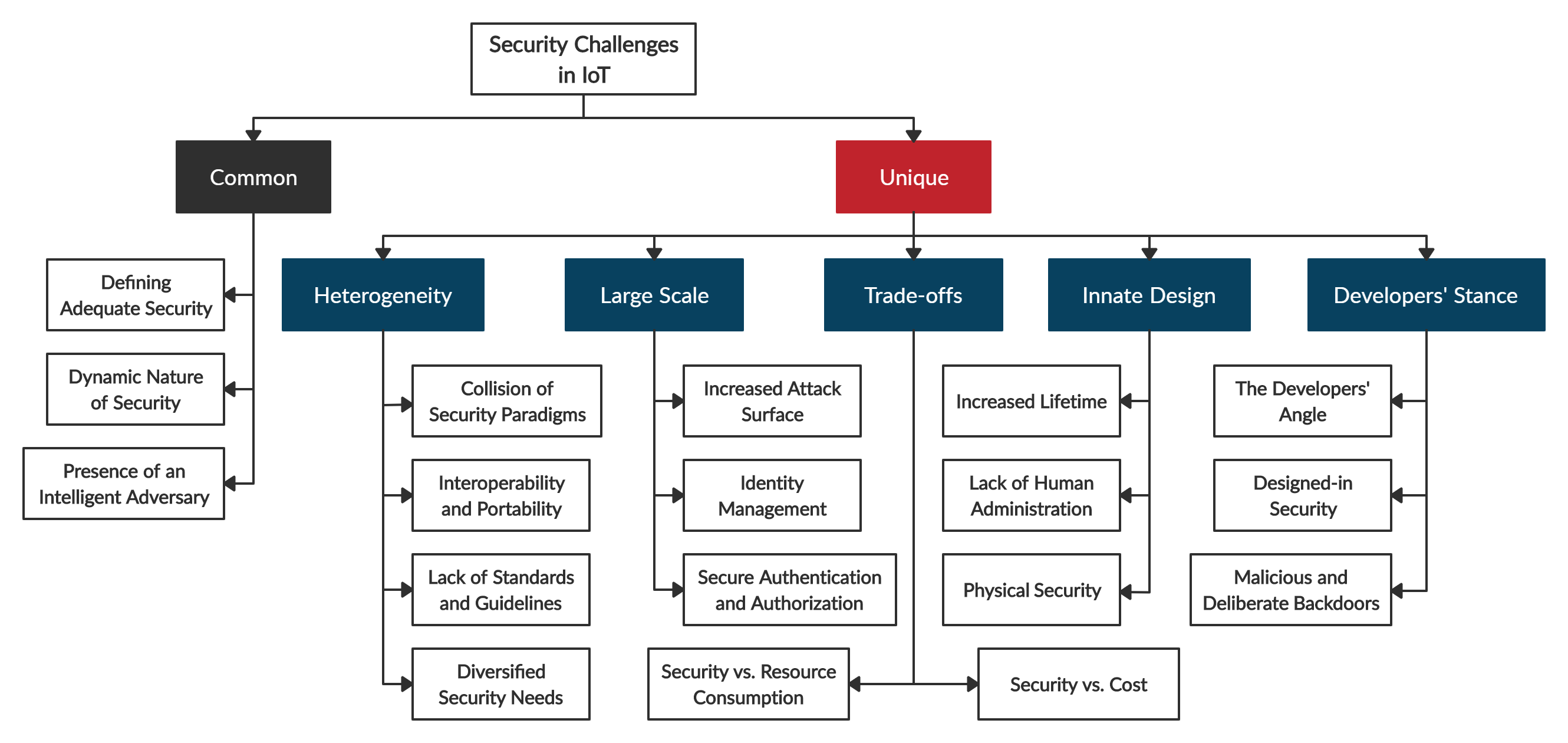}
    \caption{Classification of Security Challenges in IoT}
    \label{fig:IoT_tree}
\end{figure*}

Security is an attribute that is quickly gaining research attention as we move towards an era of ubiquitous connectivity. As our dependence on software-intensive devices grows, the security issues related to individual devices and connected systems are becoming ever more prominent. While a need to address these issues has existed from long ago, several reasons inhibit us from achieving infallible security. This section explores the common and unique security challenges related to IoT development and presents a classification of the challenges, as illustrated in figure \ref{fig:IoT_tree}.

\subsection{Common Security Challenges}
First, we take a look at the security challenges that we must face regardless of the type of system to better understand the fundamental obstacles in the path of achieving definitive security.

\subsubsection{\textbf{Defining Adequate Security:}}
For one, security is a property that is very hard to define and measure in a system. Security is not compositional, meaning a system built from secure components is not necessarily secure itself. Therefore, even if we could prove that all the devices in an IoT system are individually secure, we cannot guarantee that the system is impervious to attacks. Although there exist standard security properties of a system (i.e., the CIA triad), to account for all the real-world heterogeneous systems, authentication and accountability had to be added later to form the extended CIA model. Even then, impenetrable security remains an intractable goal, and what constitutes adequate security becomes a subjective question defined heavily by the type of system and the system stakeholder's security requirements.

\subsubsection{\textbf{Dynamic Nature of Security:}}
Another challenging aspect is the fact that security is dynamic over the lifetime of a product, and security controls must be implemented in every stage of the product lifecycle. Any component or sub-system, previously determined to not need any protective mechanism, can be deemed to be security-critical after the discovery of an exploit. This is especially concerning in emerging technologies dealing with enhanced connectivity like IoT. While the devices hitherto without connectivity features get increasingly connected to the internet, new vulnerabilities ensue. A single insecure device can then open a loophole in the system and leave the entire network exposed to an adversary. In the past few years, DDoS attacks on the IoT, such as the Mirai, Hajime, and BrickerBot attacks have proven just how severe the consequences can turn out to be \cite{DDoS_Mirai}. The integration of novel and upcoming technologies often introduces modifications to the system behavior and necessitates redefining the security requirements and revisiting the security controls already in place.

\subsubsection{\textbf{Presence of an Intelligent Adversary:}}
What makes security different from safety issues is the presence of a group of adversaries constantly trying to compromise the system through existing and undiscovered vulnerabilities. This new factor introduces the unpredictability and adaptability of human behavior and precludes the use of formal methods such as probabilistic measures to specify and verify system security. While a security engineer must constantly monitor and provide defensive mechanisms for every single vulnerability in a system, an attacker needs to find only a single vulnerability to disrupt the safe and effective operation of that system. The task gets even more challenging in IoT as the defenders not only need to defend from existing and undiscovered attacks but also attacks that will inevitably emerge with the integration of every novel IoT technology. As these adversaries become more proficient and gain access to tools and attacks with increased magnitude and sophistication, the need to secure every embedded device assisting our everyday life becomes a priority.
\thispagestyle{empty}

\subsection{Complex Trade-offs}
We then examine the security challenges that are exclusive to IoT systems and focus our discussion on challenges stemming from the heavily resource-constrained nature of devices.

\subsubsection{\textbf{Security vs. Resource Consumption:}}
Many IoT devices, especially purpose-specific sensors and actuators, are extremely constrained in terms of processing power, memory, and energy consumption \cite{RIOT}. While there are strong and robust cryptographic algorithms available, they also strain the already limited resources on these devices. Public-key operations are typically resource-intensive, and many IoT devices may not have enough memory to store a certificate or perform a cryptographic operation to validate it \cite{Denmark_Sec_Framework}. This can result in degraded performance, and thus higher latency in safety-critical applications such as wireless pacemakers and advanced driver assist systems (ADAS) in autonomous vehicles, leading to bodily harm or even potential fatalities.

Some IoT devices are designed to operate for years and may be located deep within the fabric of a system without any physical access to them. The battery capacity is directly linked to the number of computations performed during its lifetime. These devices may be involved in the critical operation of a system and are expected to perform their intended functions for a fixed period of time. Implementing rigorous security measures can shorten the expected battery life of these devices and create unwanted disruptions in the safe and reliable operation of the system.

\subsubsection{\textbf{Security vs. Cost:}}
There is a cost associated with every available security mechanism. Many designers tend to focus on the functionalities, and many companies are only interested in short-term profits \cite{Denmark_Sec_Framework}. Many developers disregard security altogether, and others develop devices with cost-effective security measures with little to no effective security in place. Some issues are also related to the consumers' priorities and their lack of concern about personal privacy. Even when some security measures are offered as a cost option, the buyers of the product often tend to go for the cheapest option with no built-in security.

\subsection{Challenges due to the large scale of devices}
With the possibility of hundreds of billions of devices being connected to the internet arises the issue of managing the scalability of the IoT ecosystem \cite{Cisco_white_paper}. The following are a few security issues that arise from a large number of devices and their increased connectivity.

\thispagestyle{empty}
\subsubsection{\textbf{Increased Attack Surface:}}

As IoT architectures are equi- pped with more and more embedded devices and network connectivities, the attack surface continues to expand. This aggregation of network-connected devices so far have led to the development of a range of hardware and communication technologies, and consequently varying vulnerabilities resulting from both. Every single exposed node in the network can act as an entry point for attackers during communication, posing the risk of data interception and modification during unencrypted transmission. A single compromised device can lead to vulnerability propagation through other linked components of the network, and the entire network may end up being compromised. As an added problem, many of these devices are from third party manufactures and involve ensuring and maintaining security throughout the entirety of the supply chain.
\hfill \break

\subsubsection{\textbf{Secure Authentication and Authorization:}}
A secure IoT system must validate the identity of the user before allowing them to access the system. In the IoT architecture, especially at the application layer, the numerous entities participating in data exchange make identity authentication and trust management a complicated task \cite{Taxonomy_IoT_Rizvy}. Issuing individual certificates for each object in IoT may be infeasible, and the absence of a global root certificate authority (CA) hinders designing an authentication system for IoT \cite{Objects_Taiwan}. Ad-hoc networks such as the mobile ad-hoc networks (MANETs) and vehicular ad-hoc networks (VANETs) present the authentication problem of entities quickly entering and leaving the network. 

Furthermore, the presence of a wide variety of applications and devices renders the creation of an exhaustive security policy and managing access permissions difficult. For example, due to the high data integrity and authenticity concerns, strong cyrptographic mechanisms such as public key cryptosystems are desirable. However, they can lead to significant computational overhead and mechanisms using cyrptographically pre-shared keys are not applicable as the rapidly growing of the number devices can overwhelm key management \cite{Objects_Taiwan}.

\subsubsection{\textbf{Identity Management:}}
Before ensuring security for individual devices, we need to deal with the issue of uniquely identifying an end-point in a scalable manner. As devices will be involved in information exchange with consumers and controllers concurrently, establishing appropriate identity controls and trust relationships between entities is crucial to maintain data privacy and exclusivity \cite{Cisco_white_paper}. Traditional naming systems for uniquely identifying a host such as the Domain Name System (DNS) is insecure and remains vulnerable to attacks such as DNS cache poisoning or man-in-the-middle attacks \cite{Objects_Taiwan}. Although DNSSEC, the security extension of DNS, can provide integrity and authentication security, the high computation and communication overhead makes it unsuitable for deployment in IoT.

Limitations of the IPv4 internet led to the development of IPv6 as one of the prominent enablers of IoT, which underscores the need for a transition from legacy platforms to and IP-enabled infrastructure \cite{Cisco_white_paper}. With the emergence of IPv6, each device can have its own unique ID and support auto-configuration. This gives IoT devices the ability to address other devices on the network individually and improves performance by a great margin. However, the lack of backward compatibility with IPv4 and the high-cost of changing internet service provider (ISP) infrastructure is still a challenge that we need to overcome. Additionally, the ISPs gaining greater visibility of network traffic compromises net neutrality and raises privacy concerns.

\subsection{Challenges due to the heterogeneity of devices}
The IoT is an evolving ecosystem comprised of a large number of heterogeneous devices. With its rapid growth comes the need to address the interoperability and portability challenges as well as the call for security across this wide array of devices, as discussed below.

\subsubsection{\textbf{Collision of Security Paradigms:}}
The design of a security mechanism for any system typically follows one of two well-established security paradigms. In paradigm A, we secure the product before it enters the market by implementing secure design mechanisms, security testing, certification, and licensing. This approach is suitable for products for which patching is impossible or difficult such as autonomous vehicles, thermostats, refrigerators, etc. Conversely, the goal of paradigm B is to make security agile. In this case, we put the product as fast as possible to the market and apply patches, updates, and mitigations to secure the device when needed. This method works well for devices that will continue to have new vulnerabilities, and applying patches are easy enough to do. Devices that fall into this category include smartwatches, smartphones, and laptops, among others.

When it comes to IoT devices, we're starting to see a collision between the two paradigms. Novel smart devices that fall into paradigm B are being increasingly connected to devices or infrastructures that must be secured under paradigm A. A typical example is the advanced metering infrastructure (AMI), where smart meters (paradigm B) are connected in two-way communication to legacy systems (paradigm A) such as electric grids or windmills to create a smart grid. This connection inherits the vulnerabilities resulting from both types of systems and complicates the system design and how security should be provided for the entire system. The way to come up with a hybrid system to manage a compromise between both paradigms is still unclear.

\subsubsection{\textbf{Interoperability and Portability:}}
Ensuring cyber and ph- ysical security for devices entails dealing with the heterogeneous hardware in IoT. WSNs, for example, can present combinations of heterogeneous sensors and actuators with general-purpose computing elements \cite{TinySec}. This creates interoperability issues and the need for easier porting of applications across different hardware. However, many current IoT OSs are designed with specific hardware in mind. A high emphasis on security may affect creating interoperable operating systems. For example, while Mbed OS from Arm provides high end-to-end security, it only supports only a small number of platforms (5 so far) \cite{HAL_IoT_OS_survey}. This necessitates the development of an IoT operating system (OS) that can provide high-level application programming interfaces (APIs) to remove hardware dependency and support interoperable security protocols.

\subsubsection{\textbf{Lack of Standards and Guidelines:}}
To develop IoT architectures with an emphasis on security, we need properly formulated, implemented, and enforced security requirements and policies throughout their life-cycle \cite{Denmark_Sec_Framework}. This task of creating a universal standard is challenging because these standards would need to consolidate security policies and requirements for the rapid growth of IoT devices with a high degree of variance. The fact that the standards would need to evolve depending on the current and emerging needs of the stakeholders such as the government, industry and users necessitates frequent revision of the requirements.

While various standards for resource-constrained IoT devices on the network level, such as those standardized by IETF (i.e., Bluetooth Low Energy (BLE), IPv6 over Low -Power Wireless Personal Area Networks (6LoWPAN)) are available, there is still a need for standards and guidelines for regulating the development of IoT devices. Compliance with the IEC 61508 CMV: Functional Safety of Electrical/Electronic/Programmable Electronic Safety-related Systems Standard\footnote{https://webstore.iec.ch/publication/22273} provides the assurance that the system provides necessary risk-reduction required to achieve adequate security for its safety-related functions. A similar standard addressing the cybersecurity requirements and relevant set of guidelines for developing IoT devices is crucial for the development of a security-centric mindset for system-designers, product developers, and consumers alike.

\subsubsection{\textbf{Diversified security needs:}}
Iot is a massive collection of devices that vary in size, application, hardware, suppliers, and resources available to those devices. The problem with providing security to all these different types of devices is the fact that security is subject to the type of device or system and varies radically from application to application. One of the prominent challenges would be the establishment of an architecture capable of handling the scalability of billions of IoT devices with varying trust relationships in the fabric \cite{Cisco_white_paper}. As mentioned before, a single security paradigm isn't enough to deal with this heterogeneity. The secure development of an IoT architecture thus needs to account for the heterogeneity of all the devices, secure communication between those devices, and the control and management of firmware updates necessary to enable that.
\thispagestyle{empty}

\subsection{Challenges due to Innate Design}
The inherent design of IoT devices, as well as the automated environment, leads to both cyber and physical security challenges in the IoT development. Some of these challenges are described below.

\subsubsection{\textbf{Increased Lifetime of Devices:}}
Cryptographic algorithms have a limited lifetime before they are either broken or need to be reworked \cite{Cisco_white_paper}. Many IoT devices are designed for increased lifetime and can continue to operate for years (e.g., smart meters can continue to function for more than 40 years) on a single battery. These long-lasting devices may outlive the lifetime of cryptographic algorithms, leading to obsolete security measures. Devices may then carry undetected malware that can persist for ages before contaminating other vulnerable devices connected to the device \cite{Objects_Taiwan}.

\subsubsection{\textbf{Lack of Human Administration:}}
Most of the IoT security mechanisms are automated and require little to no human involvement \cite{hitachi_white_paper}. This lack of human administration makes detection of undiscovered vulnerabilities improbable. Unlike IT systems, vulnerabilities in IoT systems can take several months before it can be discovered and cause both primary damage (direct damage from the incident occurrence) and secondary damage (damage as it spreads through the system). We also may not have any physical access to some of the embedded IoT devices hidden deep within the fabric of the system. That means we might not have the physical means to stop or perform a factory reset in case of a malfunction or a compromise. Therefore, developers must provide comprehensive built-in security mechanisms and a procedure for sending firmware updates to all the vulnerable models. Devices without an update mechanism must protect the initial configuration from tampering, theft, and other forms of compromise throughout the lifetime of the product \cite{Cisco_white_paper}.

\subsubsection{\textbf{Physical Security of Devices:}}
In addition to lacking proper cyber protection, a wide variety of IoT devices, e.g., security cameras and doorbells, are located in places that lack physical protection. There is a possibility of intentional or inadvertent displacement of fixed IoT devices and the theft of mobile devices\cite{Cisco_white_paper}. These devices are extremely vulnerable to being controlled or replaced with spoofed devices by adversaries, which can then be used to perform reconnaissance and further attacks. To illustrate, the OBD-II port in an autonomous vehicle, typically used to generate diagnostic information, is also accessible and exploitable by an attacker. This situation asserts the fact that we must always maintain complete control over all the embedded devices in an IoT system and the connected infrastructure, which is a challenging task.

\subsection{Challenges due to the Developers' Stance}
The following discussion explores some of the security issues from the developers' angle, their priorities, and the limited opportunities presented to them.

\subsubsection{\textbf{The Developers' Angle:}}
Some security vulnerabilities also relate to the developers of IoT devices. Device manufacturers are strongly incentivized to produce and get their products to the market as fast and cheaply as possible. Many third-party developers view security as an optional addition, and thus security measures are the first things to get abandoned in times of budget constraints and tight schedules. This results in chips being developed with a focus on functionality and little attention to security mechanisms or application of security patches. Once a product enters the market, the end-users may have little to no means of patching the device. The lack of security concerns and knowledge of a typical user, along with the carelessness and error-prone nature of humans, exacerbates this situation and leaves millions of active internet-connected devices vulnerable to attacks.

Most of the developers are not security experts, and thus their designs tend to have less focus on the security aspect. Although some developers understand the necessity and actively try to ensure security for their devices, the inherent complexity of the vast IoT network, coupled with inadequate documentation of existing IoT vulnerabilities, thwart their efforts. The fact that the implemented security mechanisms need to be maintained throughout the lifetime of the product creates reluctance and ultimately dissuades many developers from putting any security measures at all in their devices. 

\subsubsection{\textbf{Designed-in Security:}}
Many IoT devices are built with only its features and connectivity in mind. To build an adequately secure system, security measures, monitoring, and recovery mechanisms have to be built-in and continued throughout the lifetime of the product. This requirement asserts the fact that not only we have to develop individually secure IoT devices, we must also ensure the security of existing networks and the connected infrastructure throughout the development process \cite{Denmark_Sec_Framework}. 

Designed-in security also has to start from the inception of development. If security controls are not built parallelly from the earliest phase of the development lifecycle, it becomes very complex to implement it mid-way or after development. This is especially true for the ever-changing IoT ecosystem, where the integration of novel technologies will mandate modifications in the system specification. These requirement specifications can be changed during the development process, but changing an entire system usually involves incurring substantial costs.

\subsubsection{\textbf{Malicious and Deliberate Backdoors:}}
IoT devices typically lack security mechanisms such as an intrusion detection system (IDS) or antivirus to detect and remove software vulnerabilities. The reason behind this is these mechanisms require real-time scanning that can result in unaffordable overhead in resource-constrained IoT devices \cite{Objects_Taiwan}. Attackers can exploit this weakness to plant a backdoor in a vulnerable device to monitor and manipulate its operation. Additionally, many vendors intentionally insert backdoors in their software products for purposes of collecting usage information, management, and testing. However, an intelligent adversary can examine the code and employ reverse engineering technologies to discover the backdoor leading to theft and misuse of private data.

\thispagestyle{empty}

\section{Role of an Operating System}
At the heart of resolving many of these security issues lies the secure development of an OS designed specifically with IoT resource-constraints in mind. These requirements entail the development of a lightweight OS capable of running with low-energy consumption and very small memory footprints while still providing rich abstractions for execution environments \cite{Contiki_Dunkels}. The source code of the OS has to be small enough to fit both the memory and RAM usage requirements while still allowing room for applications to run on top of the system. The OS should be able to take full advantage of the power-saving modes available to the device applications and network protocols in order to conserve energy.

Although the OS cannot stop the rapid growth of IoT devices, it can streamline and direct an OS-centered development of products by providing a suitable unified platform. To accommodate the manifold devices and their requirements, the OS can provide software primitives to enable simple hardware-independent application development and APIs to allow ease of integration with the wide variety of IoT hardware and use cases \cite{HAL_IoT_OS_survey}. The issue of portability to new platforms and the need to account for various peripheral devices mandate the need for a flexible hardware model and potentially a hardware abstraction layer. The OS can tackle communication interoperability issues by supporting multiple network protocols such as Zigbee, IEEE 802.11, BLE, 6LoWPAN, 3G, 4G, etc. Software can be run on the cloud due to resource constraints, and the OS can collaborate and manage the software execution.

In IoT, diversified hardware platforms and customized operating systems make the generation of security awareness and development of a unified security solution difficult \cite{Objects_Taiwan}. Once again, an OS can be a secure platform that can facilitate and guide the development of secure products. A unified OS may provide security both on the hardware and software levels by providing features such as secure storage, secure boot, and a secure execution environment (SEE), among others. Secure transmission of data can be ensured by the inclusion of secure communication protocols through an API. Additionally, it can simplify the job of the developers by providing support for standard programming languages and mature software development kits (SDKs) to allow easier application writing, testing, and verification. The OS can implement firmware updates in the device and manage the signing, validation, and revocation on the cloud side.

We need to see IoT OS and device development from a different perspective as IoT development migrates and gets increasingly involved in safety-critical services or infrastructures. As an example, the field of automotive vehicle industry goes through rigorous checking, verification, and validations because of their safety-critical nature. As people get more concerned about the large-scale deployment of IoT, individual hardware and software would need to meet higher engineering requirements before it can be released to the market. This entails the development of IoT focused standards and guidelines that can be centered around the development of a unified OS and the features that it provides.

As security is dynamic over the lifetime of a product, security has to be designed-in the OS, and continuous enhancement of the security features through timely updates is desired. However, product designers are presented with various OSs and hardware architectures to choose from, often too many. The limited availability of information can often create confusion and overwhelm the developer, leading to a subpar choice. If there was a properly-documented definitive OS, certified by international standards to be safe and secure, developers could reliably choose that OS to leverage the in-built protection mechanisms while implementing their own and applicable third-party security on top of it. This results in a small, but more focused development of a unified platform, much like Linux, that can ultimately fulfill the needs of both the IoT industry and individual developers alike.

Our ultimate vision is to architect a highly secure platform from the ground up. To this end, we aim to direct the secure development of a unifying OS that can support the needs of the existing and emerging IoT devices while providing built-in measures to tackle the security challenges in the IoT domain.


\section{Security Requirements}
In order to build an IoT OS capable of addressing the security challenges discussed above, we first need to define the security requirements to be considered during the development phase. These requirements can vary heavily depending on the system stakeholders' needs, concerns, objectives, and system architecture under consideration. In light of the secure design principles, as delineated in the US National Institute of Standards and Technology (NIST) Special Publication 800-160 (\cite{NIST_Design_Principles}), and some effective approaches adopted by some IoT OSs, we specify the key security requirements to direct the development of a unified IoT OS below. 
\thispagestyle{empty}

\subsection{Open Source}
This requirement follows directly from the secure design principle of "Open Design". The principle emphasizes the transparency of the security mechanism and states that the security of a system should not depend on the secrecy of its design or implementation. This contradicts the idea of "Security through obscurity" where the strength of the security depends on the user's ignorance of how the security mechanism works. IoT, as an emerging technology, will include a wide variety of knowledgeable users and attract attention from resourceful attackers. If these users or attackers can grasp the mechanism's workings, they can crack the security of the system. This has been apparent in Microsoft's proprietory Windows OSs and their numerous vulnerabilities.

This leads us to the resolution that the development of an adequately secure OS should follow an open-source design. The license of the OS source code should be either copyleft (e.g., Linux) or permissive (e.g., under BSD, MIT license) with a high degree of freedom. This approach can boost the number of contributors and reviewers, resulting in better bug-fixing, and thus a secure and higher quality code \cite{HAL_IoT_OS_survey}. Copyleft licenses, in particular, can facilitate the development of an integrative community around a common code base, and ultimately balance the protection of end-users while supporting industry, as exemplified by Linux \cite{RIOT_2018_RR}.

\subsection{Isolation of Resources}
In a typical flat security model, a single vulnerability can expose the whole system to further exploitation. This model results in a fairly large attack surface, as an attack affecting any part of the OS can further compromise other parts, potentially taking control of the entire device. To reduce this attack surface, we need to completely isolate the OS resources into sections that contain the security-sensitive resources and sections that do not. This approach is in accordance with the design principle of "Isolation" and can play a major role in endorsing the segregated design of an IoT OS.

The private or critical resources segment can be comprised of the resources and functions that are security-critical, such as the secure storage (contains cryptographic keys and credentials), critical code sections, firmware updates, etc. The other segment, namely the public or uncritical resources segment, should then contain resources and functions that are not directly related to the security of the OS. Examples of these are network stacks, application protocols, device management, and diagnostics, etc. This makes the attack surface as small as possible and restricts the potential damage done once a vulnerability has been exploited. 

The two isolated regions should be different in how they access the memory and address program execution. For one, the public side should never be able to write directly to flash memory. The OS should provide APIs that allow operations to the private segment, but the critical resources should never be directly accessible to the public segment. Cryptographic keys, for example, should never be allowed to leak into the public segment. Additionally, the private side should provide a secure runtime environment for secure code execution and leave small memory footprints. Finally, the private side can verify the integrity of device data and perform a clean reset if any sign of compromise has been found. In this way, even if an attack on the public side is successful, the security-sensitive data remain safe.

\thispagestyle{empty}
\subsection{Customized Development Approach}
The isolation of the public and critical resources discussed in the last section allows us to use a customized development approach tailored to the needs of each section. In light of the rapid growth of IoT devices, the public section would need much faster development. With the target of getting devices to the market as early as possible, we must allow developers to adopt an iterative approach and quicker product development and delivery. While this approach can potentially open the systems to vulnerabilities, attacks on the public side cannot affect the private side of the OS, keeping the system secure. This meets the faster development requirement for IoT without compromising the security of the device.

The goal for the private side, however, is the protection of critical resources and the steady, measured, and exhaustive development of security mechanisms. As code with a certain degree of complexity can rarely be expected to be bug-free, the development of this section will need to account for constant testing, reviewing, bug fixing of existing code, and patching vulnerabilities resulting from new features. As security issues become more debated, compliance with updated standards and guidelines will also be an issue. All these requirements mandate that the development of the code on the private section be slow, stable, and rather unchanging, ultimately resulting in a mature and high-quality codebase through constant review and revision.


\subsection{Hybrid Programming Model}
The programming model in an OS defines the way an application developer can model the programs for that OS. Typical programming models used for IoT OSs are event-driven systems and multi-threaded systems \cite{HAL_IoT_OS_survey}. In an event-driven system, each task detects and reacts to triggers by an external event, such as an interrupt or stimulus. In contrast, a multi-threaded system can allow multiple threads of concurrent execution of a program. Each task in this model can run in its own thread context, and an inter-process communication (IPC) API allows communication between the tasks. Each thread requires its own stack, and the stack typically has to be over-provisioned, consuming memory resources \cite{Contiki_Dunkels}. Purely from a performance perspective, event-driven systems can share the limited resources between all processes (by using the same stack) and never run into concurrency problems, as illustrated by their use in multiple WSN OSs, including Contiki and TinyOS. However, more complex and novel IoT applications may require capabilities of a fully-fledged OS to ease application design, i.e., multi-threading  \cite{RIOT}.

From a security viewpoint, however, multi-threading poses some serious issues. First of all, it introduces increased complexity and requires expertise from the application developers making error identification and testing much more complex. Locking mechanisms are needed to prevent concurrently running threads from modifying shared resources, and mishandled concurrency can lead to unfamiliar or unwanted behavior. On the other hand, event-driven systems present a simple design and easier testing and verification of the model. Contiki is based on an event-driven model but still provides optional preemptive multi-threading support for programs with explicit needs \cite{Contiki_Dunkels}. To achieve maximum security while leveraging the benefits of multi-threading, a hybrid model similar to that of Contiki may be adopted.

\subsection{Simplification of Security Choices}
From the secure design principle of "Economy of Mechanism" follows the idea of streamlining the development of IoT. Adherence to this principle means architecting the OS in a way that simplifies the job of the developers rather than letting each developer come up with a security mechanism for themselves. For example, we can limit the number of security choices offered by the OS architecture to simplify the design, allowing for simpler and more rigorous testing. The goal is to support the limited security knowledge of the developers by providing them with simplified design choices and minimal, but critical information. The reduction of options will result in fewer assumptions made by developers, ultimately leading to an effective security architecture with fewer risks. With the vision of building a secure and better platform from the ground up, this meets the need to support the numerous developers involved in key roles in the development of the IoT industry.

\subsection{Standard Programming Language and Code Maturity}
In designing an OS for IoT devices, we typically have the choice to select (i) a standard programming language (typically ANSI C or C++), and (ii) an OS-specific language or dialect \cite{HAL_IoT_OS_survey}. Selection of a dialect, such as nesC (a dialect of C) used by TinyOS, requires a higher learning curve but can support system performance and safety through enhancements missing in low-level languages like C. Standard programming languages, on the other hand, can allow for a near-zero learning curve and support the use of well-documented and mature tools. This increases code reviewability and allows simpler debugging following the two design principles of open design and a simplified mechanism. The maturity of the code - typically defined by the age, documentation per lines of code, and the size of the development community - can also be a good indicator of high-quality and secure implementation of the OS.

\subsection{Targeted Mediation}
The design principle of "Complete Mediation" states that every access to the operations of the device must be checked against the access control mechanism. To fully implement complete mediation, the OS should limit the caching of information and check the access control policy every time an entity requests access to system resources. While this method strongly augments security, this is very resource-intensive. An OS for resource-starved IoT devices should thus take a different approach to provide mediation. Rather than providing full mediation for every single access, the OS should provide mediation for all security-sensitive operations. This will allow the OS to efficiently perform public operations in real-time with relatively no performance overhead while still providing adequate security for operations on critical system resources. This approach requires redefining the security requirements and revising the access control policies.

\subsection{Compliance with IoT-centric Standards and Protocols}
The distinct needs of application development and IoT device constraints lead us to the resolution that a single OS may not be capable of fulfilling all of these requirements. This foregrounds the need for standardized protocols to improve interoperability and portability of applications across varying hardware and platforms. RIOT, for one, uses a POSIX like API to achieve interoperability for all of its supported hardware (from 16-bit microcontrollers to 32-bit processors) and is aiming towards achieving full POSIX compliance \cite{RIOT}. However, compliance with traditional standardized APIs such as full POSIX compliance can only be achieved by a few OSs in PC and may not be suitable for IoT OS development \cite{HAL_IoT_OS_survey}. 

On the other hand, we also need standard interoperable security protocols for the network layer. Datagram Transport Layer Security (DTLS), standardized in 2006 by IETF to reduce power consumption and delays while maintaining similar security protections as TLS, can be a key enabler of secure IoT connectivity. While work focused on standardization has indeed progressed over the past few years, many works are either incomplete or are ahead of product implementation, resulting in severe complexity in the integration of security libraries into developers' hardware of choice \cite{IETF_standards}. Therefore, the development of developer-friendly IoT-centric open standards both on the system level and the network level is imperative for stable and secure IoT OS development.

\thispagestyle{empty}
\subsection{Support for Lightweight Cryptography}
In light of the constraints of IoT devices, typical cryptographic mechanisms can prove to be too resource-intensive and expensive \cite{Cisco_white_paper}. Therefore, there is a need to implement a new optimized and lightweight cryptographic scheme designed specifically for resource-constrained embedded devices. This cryptographic scheme would need to ensure a high level of security while using minimal memory, power, and execution speed requirements \cite{Denmark_Sec_Framework}. Elliptic-curve cryptography (ECC) uses much smaller key sizes compared to non-EC cryptography to provide equivalent security, making it a suitable scheme for embedded IoT devices. The Cryptographic API of the OS should support security services for signing, validation, and revocation of existing and emerging lightweight cryptographic schemes.




\subsection{Support for Common Security Mechanisms}
With strict IoT constraints in mind, the OS should provide as many common security mechanisms available for traditional OSs as possible. First of all, the OS should provide secure storage to enable the physical protection of cryptographic keys and other security-sensitive data. The secure storage should also be persistent (e.g., on-chip ROM memory) over the lifetime of the product to prevent loss of data during power cycles \cite{Denmark_Sec_Framework}. Various debugging and diagnostic interfaces, e.g., the JTAG interface used to debug errors during manufacturing and development, are potentially exploitable and must be secured.

The OS can also adopt a secure booting mechanism to bring the device to a familiar and trusted state after system startup. This is similar to the verified boot\footnote{https://source.android.com/security/verifiedboot} feature in Android, where a full chain of trust is established during device boot up, and integrity and authenticity of the next stage is verified by each stage before execution is handed over. As discussed before, an isolated SSE can provide additional security by running any operation requiring security-sensitive data in a trusted environment. As long as the computation overhead is small enough, security modules, such as MiniSec for TinyOS, can also be implemented on top of the OS architecture to achieve targeted protection \cite{MiniSec}. During firmware updates, rollback protection should be enabled to stop devices from downgrading to older versions and preclude the possibility of persisting vulnerabilities.
\thispagestyle{empty}

\section{Modern IoT OSes}
In this section, we examine some of the prominent IoT OSs and compare and contrast their approaches to architecting an OS to the security requirements discussed in the previous section. We also briefly explore some of the OS-specific in-built security approaches and security evaluations performed in the literature.

\subsection{Contiki}
Contiki was originally developed in 2002 with the goal of supporting WSNs running constrained 8-bit MCUs \cite{HAL_IoT_OS_survey}. Contiki is open source and is available under a BSD license on GitHub\footnote{https://github.com/contiki-os/contiki} and various other platforms. Development of Contiki has focused on power efficiency and lightweight memory management, typically requiring only 2 kilobytes of RAM and 60 kilobytes of ROM for configuration \cite{Contiki_Dunkels}. Contiki supports the dynamic loading and unloading of individual applications or services at runtime. As individual application binaries are smaller than the entire system binary image, it allows Contiki to use less energy and less time for transmission of the binary image through the network. This also allows multiplexing of hardware of a sensor network across multiple applications or users.

Contiki has received frequent research attention from academic users, and over the years has turned into one of the most widely used OSs for WSN capable of running 16-bit and modern ARM 32-bit MCUs. Contiki uses a hybrid programming model. The base system runs on an event-driven kernel, and preemptive multi-threading is implemented as an application library on top of that \cite{Contiki_Dunkels}. The choice of using the multi-threaded model is optional, and applications that specifically require such a model of operation can be linked to the library. Contiki is written primarily in C, while still providing runtime environments for implementation in Java and Python. A large variety of independently developed forks exist, including many that are closed source, resulting in both highly-mature and experimental codebase sections.

Contiki uses the Cooja/MSPsim simulator to support debugging and various testing mechanisms, including unit testing, regression testing, and full-system integration testing \cite{HAL_IoT_OS_survey}. It also provides features such as a shell, a file system, a database management system, cryptographic libraries, among others. Additionally, Contiki supports many lightweight communication standards, including IEEE 802.15.4, 6LoWPAN, CoAP, MQTT, TSCH, and RPL. Its core IPv6 has been certified with a silver certification in the IPv6 Ready Logo Program \cite{HAL_IoT_OS_survey}. Finally, the core system provides a basic abstraction, featuring a hardware-independent software infrastructure for the application-driven nature of the heterogeneous sensor devices.

McBride et al. in \cite{EWSN_SecAnalContiki} perform a security analysis of Contiki using static program analysis tools. In static analysis, the code is interpreted without execution for errors to identify evasive bugs and unwanted program behavior. The analysis shows that although the number of potential bugs has increased across different Contiki releases, the average bug density (number of bugs per thousand lines of code) has consistently decreased over time. They also identify two major vulnerabilities (a use-after-free vulnerability and a persistent cross-site scripting (XSS) attack) and a few minor issues, leading to their documentation and patching.

The authors of \cite{ContikiSec} examine the performance characteristics of different security primitives, such as block ciphers, cipher-based message authentication code (CMAC), etc. under Contiki. Additionally, they present ContikiSec, a secure network layer for Contiki, designed for secure transmission over wireless sensor networks. ContikiSec supports a configurable design centered around three security modes that focus on preserving confidentiality, integrity, and authentication. The design aims to provide additional security while balancing low-energy consumption and small memory footprints with the strongest security mode (ContikiSec-AE) consuming approximately 15\% more energy compared to Contiki running in default mode.

\subsection{TinyOS}
TinyOS, developed since 2000, is one of the first and most widely used OSs in WSNs. The core OS can fit within a memory of only 400 bytes, and many applications require only 16 KB of memory, capable of supporting the efficient, low-power operation of complex, concurrent programs \cite{levis_TinyOS}. The OS is implemented in a C dialect called nesC, which puts some restrictions on C (e.g., use of function pointers) to improve code efficiency and robustness.  The source code of TinyOS is available under the BSD license on GitHub\footnote{ https://github.com/tinyos/tinyos-main}, the complex and customized nature of which precludes the formation of bigger community around TinyOS \cite{HAL_IoT_OS_survey}. TinyOS follows a purely event-driven model, supporting modules of components with a high level of sophistication and optimization; optimized code can be faster and smaller than even original hand-written code in C \cite{levis_TinyOS}.

\thispagestyle{empty}
NesC provides automated static race detection, removing the concern about bugs generated due to concurrency during program composition \cite{levis_TinyOS}. Various simulators in TinyOS, e.g., TOSSIM, Viptos, QualNet, Avrora, EmTOS, etc. can be used to simulate different implementations and applications of TinyOS. TOSSIM, in particular, can be used to analyze TinyOS at a very basic level and find numerous bugs in its source code and its various applications \cite{TinyOS}. TinyOS also facilitates the security assessment of WSNs by allowing the study of common attacks (e.g., wireless injection attacks, DOS attacks, man-in-the-middle attacks) on TinyOS.

Authors in \cite{TinySec} present TinySec, a link-layer cryptographic solution for TinyOS, designed to fit the resource constraints and security needs of IoT. TinySec has very low impacts on bandwidth and latency and features minimal energy overhead as different modes of security require anywhere from 3\% (TinySec-Auth, least secure mode) to 10\% (TinySec-AE, most secure option) overhead in sensor network applications. The block ciphers used in TinySec are RC5 and Skipjack due to their speed and suitability for software implementation on embedded microcontrollers, and TinySec provides ease of switching between both. TinySec has been widely implemented in WSN, including implementations in custom hardware, and has paved the way to the development of many other security projects such as TinyPK, TinyCrypt, and SecureSense, among others.

MiniSec is an open-source general-purpose security module designed for TinyOS to offer high-levels of security within the energy consumption and memory constraints of wireless sensor nodes \cite{MiniSec}. It was designed to leverage the low energy consumption of TinySec and high-levels of security provided by ZigBee. Two available communication modes, single-source communication, and multi-source broadcast communication allow MiniSec to achieve high-levels of data secrecy, authentication, replay protection. MiniSec-B (used for broadcast communication) always outperforms TinySec and requires as little as \(\frac{1}{3}\) of the energy consumed by TinySec.

\subsection{RIOT}
One of the newer members of the IoT OS family is RIOT, which has received significant academic research attention and grassroots community support since its emergence in 2012. Part of the reason for this popularity is RIOT's IoT-centric design and a developer-friendly API, allowing for standard programming in C or C++ and the use of well established debugging tools such as GCC, GDB, and Valgrind \cite{RIOT_2018_RR}. Applications can also be developed under Linux or Mac OS using the native port, and are highly portable, running seamlessly on a wide range of hardware, including 8-bit, 16-bit, and 32-bit platforms. RIOT is open-source, and the source code is available under GNU Lesser General Public License (LGPLv2.1) in GitHub\footnote{https://github.com/RIOT-OS/RIOT}. Its energy-efficient design requires only ~ 1.5kB min RAM and ~ 5kB min ROM for execution. RIOT has achieved partial POSIX compliance and is working towards attaining full POSIX capabilities.

To the best of our knowledge, RIOT development so far has focused more on its real-time and multi-threading support for IoT as opposed to security. Its modular microkernel structure provides minimal isolation by preventing bugs in a single component (e.g., device driver, or the file system) from harming the whole system \cite{RIOT}. RIOT provides full multi-threading support following the classical multi-threading concept with memory passing inter-process communication (IPC) between threads and minimal computational and memory overhead (\(<\)25 bytes per thread) \cite{HAL_IoT_OS_survey}. Applications can create as many threads as needed, limited only by the memory and the stack size available for each thread.

\subsection{FreeRTOS}
FreeRTOS development evolved around the goal of creating a real-time operating system (RTOS) to cater to the real-time needs of industrial and commercial contexts \cite{FreeRTOS}. Developed since 2002, FreeRTOS is available under the MIT open-source license and enjoys support from many professional and community contributors. Its tiny kernel (memory footprint can be as small as 9kB) is capable of supporting more than 40 MCU architectures, some of which also provide a tick-less power-saving mode, ensuring high portability and energy efficiency. FreeRTOS supports a multi-threaded programming model. The OS itself is written in C while providing seamless C++ application support and an integrated IDE \cite{HAL_IoT_OS_survey}. 

With no networking capabilities built-in, FreeRTOS has to depend on additional tools and libraries available to its ecosystem \cite{HAL_IoT_OS_survey}. While these resources are available, this creates a heavy reliance on third-party providers for network stacks as well as testing and debugging purposes. Detailed IoT-specific pre-configured demos and references allow developers to take advantage of libraries to establish a secure connection to the cloud. A fork of the FreeRTOS code base, SafeRTOS, is developed with significant considerations for safety and supports a wide range of international development standards.
\thispagestyle{empty}

\begin{figure}[t]
    \centering
    \includegraphics[width=0.4\textwidth]{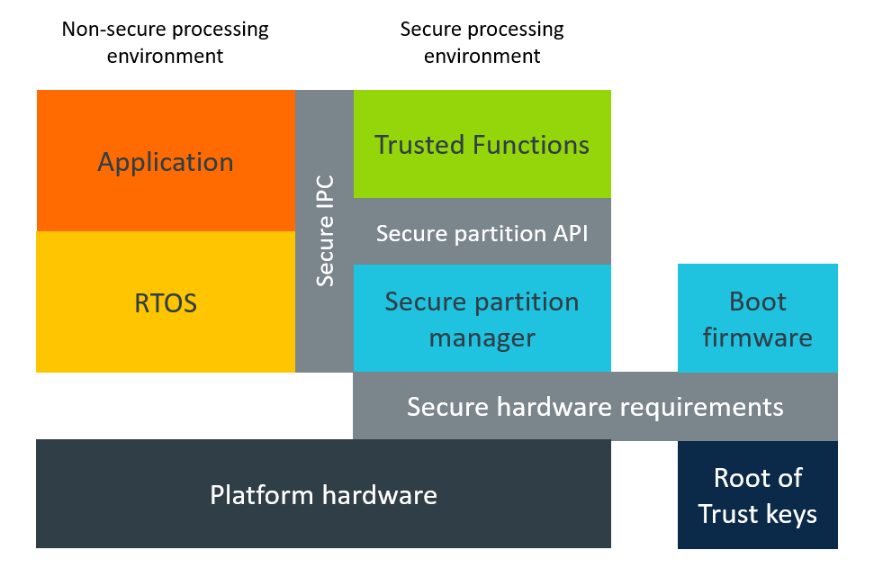}
    \caption{Platform Security Architecture (PSA) of Mbed OS}
    \label{fig:MbedOS}
\end{figure}

\subsection{Mbed OS}

Mbed OS from Arm is designed exclusively with IoT implementation in mind to facilitate the development of connected products based on the Arm Cortex-M microcontroller~\cite{Arm_Mbed_OS}. It is a free, open-source OS under the Apache~2.0 license developed primarily by Arm, alongside its partners, and numerous individual developers around the world. Mbed OS is written using the C and C++ programming languages. It features automated inclusion of modular library structures and an online integrated development environment (IDE) for ease of program development. It has a RTOS core, enabling deterministic and multi-threaded real-time software execution. Mbed OS has achieved a Thread certification and supports many lightweight communication protocols, including BLE, Thread, 6LoWPAN, Mobile IoT (LPWA), Ethernet, and WiFi.

The development of Mbed OS emphasizes high end-to-end security by implementing security mechanisms in device hardware, software, communication, and maintaining it throughout the device lifecycle. It features hardware-enforced isolated security domains at the lowest level of the OS to restrict access to memory and peripherals. The platform security architecture (PSA) implementation of Mbed OS, illustrated in figure \ref{fig:MbedOS}, shows the isolation between a secure processing environment (SPE) and a non-secure processing environment (NSPE). The SPE contains cryptographic assets, credentials, and critical code sections, and provides a SEE for the execution of security functionalities \cite{MbdeOS_PSA}. The NSPE, on the other hand, contains application firmware, OS kernel, libraries, and other nonsecure hardware resources.

The secure partition manager (SPM) is a PSA-compliant software hypervisor that manages the isolation on a hardware level and provides standardized APIs to ensure secure IPC between SPE and NSPE. Correct use of SPM provides resiliency against persisting malware and prevents secret data leak between different modules in an application. The Mbed OS API allows the simple development of portable applications while leveraging its multi-layer security and communication features. Communication security is reinforced by the simple inclusion of secure sockets layer (SSL) and transport layer security (TLS) protocols using an API.

\section{Related Work}
Zhang et al. \cite{Objects_Taiwan} define the "things" in IoT as physical or virtual objects with connectivity and discuss 7 major areas of ongoing and prospective research work. The authors primarily address the security challenges stemming from the heterogeneity and large scale of IoT devices and networks. Their research emphasizes the need for research in inheriting software vulnerabilities and malware and call for novel IoT-specific and secure object identification, authentication, and lightweight cryptographic protocols.
\par
The authors of \cite{Denmark_Sec_Framework} highlight the need for embedded security and propose an embedded security framework for the development of IoT. Their framework is based on 6 key security requirements, featuring a mix of hardware and software-level security for a hybrid, cost-effective implementation. The authors, however, acknowledge the framework's dependency on precise definitions of parameters such as resource constraints, and network and system specifications.
\par
In light of the differences between the IoT architecture and traditional IT systems, the authors of \cite{hitachi_white_paper} assert the importance of immediate detection and providing provisional measures to isolate the anomalous section and prevent the spread of damage. They propose a seven-step systematic approach, named as the cyber kill chain model, to identify and prevent misuse of built-in OS commands or software.
\par
The authors of \cite{Borgohain2015SurveyOS} perform a survey of IoT operating systems in which they focus on identifying the communication protocols and software development kits (SDK) available to each OS. Their research acknowledges the distinctions between traditional and IoT-specific operating systems and underscores the need for additional security measures to build robust WSNs.
\par
Hahm et al. \cite{HAL_IoT_OS_survey} perform a detailed analysis of the specific requirements that must be satisfied by an OS to run on low-end IoT devices. They briefly survey the applicable OSs for class 1 and class 2 devices in the IoT domain, focusing on the need to identify a unifying OS for all IoT devices. Additionally, their research specifies key design choices, focusing on distinct technical and non-technical properties concerning the development of an IoT-specific OS. Finally, they perform a detailed case study on representative OSs based on three different categories (event-based OSs, multi-threading OSs, and pure RTOSs) underscoring the resolution that different OSs fit different criteria, and a unifying architecture and capabilities of an OS are yet to be determined.
\par
In contrast, our work aims to facilitate the understanding of fundamental security challenges by exploring the various facets of the IoT ecosystem. We also present the role of an OS in addressing many of these issues and identify some key security requirements to support the steady and secure development of IoT devices. Additionally, we survey the modern literature in the IoT OS domain to determine adherence to our security requirements and discuss some security-centric approaches and evaluations.

\thispagestyle{empty}


\section{Conclusion and Future Work}
In this paper, we provide a classification of the common and unique security challenges arising from the different aspects of IoT devices and the architecture. More specifically, we focus our study on the complex trade-offs due to resource-constraints, and challenges stemming from the inherent design, large-scale, and heterogeneity of devices. We also dive into the developers' perspective to try and understand the issues that inhibit them from designing adequately secure products. To the best of our knowledge, an effort to classify all the security challenges in IoT development did not exist prior to our work.

We also discuss the pivotal role of an OS in resolving many of the aforementioned challenges and acting as a concrete platform to direct the changes necessary for a better IoT architecture. In addition, we specify security requirements following secure design principles and highlight the principles of "Open Design," "Isolation," and "Economy of Mechanism." We also suggest the use of a hybrid programming model, standard and mature programming languages and tools, and emphasize the need for IoT-specific open standard development.

Finally, we survey some of the dominant IoT OSs and compare and contrast the approach they have taken to our security requirements. We also examine some of the security-centric evaluations done on specific OSs and discuss their findings to gain greater insight into the current state of development for IoT OSs.

Our conclusion is that current IoT OS development has a divergent focus, and with the exception of Mbed OS, very few OSs are emphasizing designed-in security required to architect a secure design platform. Most OSs surveyed in this paper fail to comply with many of the specified security requirements, leading us to the resolution that the secure development of a unifying OS, around which secure development of IoT devices can be centered, is still far away.

Our future work would involve a more rigorous specification of the security requirements in compliance with published security standards and guidelines for IoT OS development. The ultimate goal is the development of a security requirement framework to define and direct IoT OS development towards a secure and resilient IoT architecture.

\thispagestyle{empty}


\bibliographystyle{ACM-Reference-Format}
\bibliography{main}
\thispagestyle{empty}

\end{document}